\documentclass{jfm}

\usepackage{newtxtext}
\usepackage{newtxmath}
\usepackage{natbib}
\usepackage{hyperref}
\hypersetup{
    colorlinks = true,
    urlcolor   = blue,
    citecolor  = black,
}

\newcommand{\RomanNumeralCaps}[1]
\linenumbers

\usepackage{epstopdf, epsfig}
\usepackage{graphicx,color,xcolor}

\definecolor{harvardcrimson}{rgb}{0.79, 0.0, 0.09}
\definecolor{venetianred}{rgb}{0.78, 0.03, 0.08}
\definecolor{limegreen}{rgb}{0.2, 0.8, 0.2}
\definecolor{candyapplered}{rgb}{1.0, 0.03, 0.0}

\providecommand\bnabla{\mathbf{\nabla}}

\newcommand{\Ro}{R_\Omega}
\newcommand{\Roc}{R^c_\Omega}
\newcommand{\Prandtl}{Pr}
\newcommand{\Nu}{Nu}
\newcommand{\J}{J}

\newcommand{\vetU}[0]{\mbox{\bf{{\em U}}}}
\newcommand{\vete}[0]{\mbox{\bf{\em{e}}}}

\title{Much faster heat/mass than momentum transport in rotating Couette flows}

\author{Geert Brethouwer
  \corresp{\email{geert@mech.kth.se}}}

\affiliation{Department of Engineering Mechanics, FLOW, KTH,
SE-100 44 Stockholm, Sweden}

\begin{document}

\maketitle

\begin{abstract}
Heat and mass transport is generally closely correlated to momentum transport
in shear flows. This so-called Reynolds analogy between advective heat or mass 
transport and momentum transport hinders efficiency 
improvements in engineering heat and mass transfer applications.
I show through direct numerical simulations that in plane Couette 
and Taylor-Couette flow
rotation can strongly influence 
wall-to-wall passive tracer transport and make it much faster 
than momentum transport, clearly in violation of the Reynolds analogy.
This difference between passive tracer transport,
representative of heat/mass transport, and momentum transport 
is observed in steady flows with large 
counter-rotating vortices at low Reynolds numbers 
as well as in fully turbulent flows at higher Reynolds numbers.
It is especially large near the neutral (Rayleigh's) stability limit.
The rotation-induced Coriolis force strongly damps the streamwise/azimuthal
velocity fluctuations when this limit is approached while tracer fluctuations
are much less affected. Accordingly, momentum transport is much more reduced
than tracer transport, showing that the Coriolis force breaks the Reynolds analogy.
At higher Reynolds numbers this strong advective transport dissimilarity is
accompanied by approximate limit cycle dynamics with 
intense low-frequency bursts of turbulence
when approaching the neutral stability limit.
The study demonstrates that simple body forces can cause clear dissimilarities 
between heat/mass and momentum transport in shear flows.
\end{abstract}


\section{Introduction}
Advective transport of heat and mass by fluid motions is fundamental
to planetary and astrophysical processes and many engineering
applications \citep{Balbus1998,Kays}.
Efficient advective transport contributes to energy savings
in buildings \citep{Lake}, process industry \citep{Keil} and data centers \citep{Alben},
and can be obtained by e.g. applying wall roughness \citep{Zhu} and
flow-control \citep{Yamamoto,Kaithakkal}.
Especially optimal transport given a minimal power input
generates energy savings in applications \citep{Alben,Motoki},
but optimization is challenging since flow vortices and eddies 
generally transport momentum and heat/mass at similar rates.
This so-called Reynolds analogy between 
transport of momentum and heat/mass 
was postulated by \cite{Reynolds}, 
and applies to many shear 
flows \citep{Kays1994,Kays,Pirozzoli} including astrophysical flows \citep{Guan}
when Prandtl numbers are close to unity \citep{Ziefuss}. 
The Reynolds analogy is used for modelling advective transport in engineering 
\citep{Kays}, geophysical \citep{Bretherton} and astrophysical flows
\citep{Birnsteil}, but implies that higher 
heat/mass transfer goes together with higher momentum transfer 
and thus power input \citep{Yamamoto}.

In recent theoretical studies, incompressible steady flows are computed
that maximize heat transfer for a given power input \citep{Hassanzadeh,Alben,Motoki,Souza}.
\cite{Motoki} consider plane Couette flow and show that the optimized flow
has a much higher heat transfer for a given power input than the ordinary turbulent flow.
The computed optimized flows are not required to obey known
momentum equations in these theoretical studies, that is, these optimal flows 
can be obtained applying a body force, but the body force can be arbitrary and
does not (necessarily) have a familiar form. 
It is therefore not clear if this optimal transport is realizable,
although \cite{Alben} and \cite{Motoki} suggest that optimal flows can be approached
by applying smart forcing or control techniques.
I show through direct numerical simulations (DNSs) that in existing flows,
namely incompressible plane Couette flow (PCF) and Taylor-Couette flow (TCF)
subject to a Coriolis force, passive tracer transport can be much 
faster than momentum transport, in violation of the Reynolds analogy.
It is thus possible to significantly change the ratio of 
wall-to-wall heat/mass to momentum transport by a simple
Coriolis body force.
Momentum transport in Couette flows has been explored extensively
\citep{Salewski,Grossmann} owing to its relevance for e.g. astrophysics.
TCF with heat or mass transport finds applications in, for example, 
cooling of electrical motors \citep{Fenot}, and
chemical reactors and bioreactors \citep{Nemri}.

\section{Governing equations and numerical procedure}\label{num}
TCF is a shear flow created between two rotating concentric cylinders while
PCF is the small-gap limit $d/r_i \rightarrow 0$ ($\eta=r_i/r_o \rightarrow 1$)
of TCF, where $d$ is the gap between the cylinders/walls and $r_i$ and $r_o$ 
the inner and outer radius, respectively. In these flows I study
passive tracer transport mimicking heat and mass transport when 
the temperature/mass does not affect the flow.
Hereafter, the passive tracer is called temperature for convenience but
the only body force affecting the flow is the Coriolis force
that does not perform any work.

Fluid motion and passive tracer transport in the PCF and TCF DNSs
are governed by the Navier-Stokes and advection-diffusion equation
\begin{eqnarray}
\frac{\p \vetU}{\p t} + \vetU \cdot \bnabla \vetU &=&
- \bnabla P + \frac{1}{\Rey} \bnabla^2 \vetU - \Ro (\vete_z \times \vetU)\label{nse},\\
\frac{\p T}{\p t} + \vetU \cdot \bnabla T &=&
\frac{1}{\Prandtl\Rey} \bnabla^2 T
\label{te}
\end{eqnarray}
together with $\bnabla \cdot \vetU=0$.
The imposed azimuthal (streamwise) velocity 
and temperature at the inner and outer no-slip and iso-thermal walls
$\pm U_w$ and $\pm T_w$, respectively, are constant.
Velocity $\vetU$ is normalized by $U_w$, temperature $T$ by
$T_w$, and length by $d$.
The modified non-dimensional pressure $P$ includes the centrifugal force \citep{Salewski}.
The rotation axis, defined by the unit vector $\vete_z$, is the spanwise and central axis
in PCF and TCF, respectively, as in \cite{Brauckmann2016},
and is parallel with the mean flow vorticity.
Sketches of the flow geometries are presented in the supplementary material.
A Reynolds number $\Rey= \Delta U d /\nu$ 
and rotation number $\Ro = 2 \Omega d / \Delta U$, where $\Delta U= 2 U_w$,
$\nu$ the kinematic viscosity and $\Omega$ the imposed system rotation,
characterize the flow. 
$\Ro$ is defined such that it is negative for cyclonic flows 
(same sign for shear and rotation) and positive for anti-cyclonic flows. 
These parameters are equivalent to the shear Reynolds and rotation 
numbers used by \cite{Dubrulle} and \cite{Brauckmann2016}.
The rotating reference frame for TCF can naturally be translated back to a
laboratory reference frame \citep{Ezeta}.
$\Prandtl=\nu/\alpha$ is the Prandtl number with 
$\alpha$ the thermal diffusivity.

From (\ref{nse}) and (\ref{te}) follows that in PCF
the wall-to-wall mean dimensionless momentum 
$
\J^m = \langle U V \rangle - \partial_y \langle U \rangle /\Rey
$
and heat fluxes
$
\J^h = \langle V T \rangle - \partial_y \langle T \rangle / (\Rey \Prandtl )
$
are conserved, and in TCF 
the angular velocity flux
$
\J^m = r^3 \left ( \langle V \omega \rangle -  \partial_r \langle \omega \rangle \right/\Rey )
$
and heat current
$
\J^h = r \left [ \langle V T \rangle - \partial_r \langle T \rangle / (\Rey \mbox{Pr} ) \right ]
$
are conserved \citep{Brauckmann2016}.
Here, $\omega=U/r$ is the angular velocity, $U$ and $V$ the streamwise (azimuthal) and 
wall-normal (radial) velocity,
and $\langle ... \rangle$ denotes averaging over time and area at constant 
wall-normal (radial) distance in PCF (TCF).
The Nusselt numbers $\Nu_m = \J^m/\J^m_{lam}$ and $\Nu_h = \J^h/\J^h_{lam}$
quantify the wall-to-wall (angular) momentum transport \citep{Brauckmann2016}
and heat transport, respectively.
The subscript 'lam' implies the molecular (conductive) flux for laminar flow.
For laminar PCF
\(
    \J^m_{lam} = - 1 / \Rey
\)
and
\(
    \J^h_{lam} = - 1 / {\Rey \Prandtl}.
\)
For laminar TCF
\(
    \J^m_{lam} = - 2\eta/(\Rey (1-\eta)^2)
\)
\citep{Brauckmann2016}
and
\(
    \J^h_{lam} = 2/(\Rey \Prandtl\, \ln \eta).
\)
$\Nu_m$ specifies the force (torque) needed to shear the flow in 
units of that in laminar PCF (TCF), and $\Nu_h$ the heat flux
in units of that in laminar flow.

To study the influence of Coriolis forces on momentum and heat transfer 
I have carried out several DNS series at constant $\Rey$ up to $40\,000$
and varying $\Ro$, i.e., eight DNS series of PCF at a constant 
$\Rey=240$, 400, 800, 1600, 3200, 6400, $17\,200$, $40\,000$,
respectively, and eight DNS series of TCF at a constant
$\Rey=400$, 1152, 2593, 3889, 8750, $19\,688$, $29\,531$, $40\,000$,
respectively, all at varying $\Ro$.
The supplementary material presents tables with 
$\Rey$ and $\Ro$ parameters of all DNSs.
In PCF, $\Prandtl=1$ and in TCF, $\Prandtl=0.7$ and $\eta = 0.714$.

The governing equations for PCF are solved with a Fourier-Fourier-Chebyshev
code, with periodic boundary conditions in the streamwise and spanwise 
directions \citep{Chevalier}.  
The computational domain size is $6\pi d$ and $2\pi d$ in
the streamwise and spanwise direction, respectively,  which is large
enough to accommodate several pairs of counter-rotating
large-scale vortices.
The governing equations for TCF in cylindrical coordinates are solved
with a Fourier-Fourier-finite-difference code \citep{Boersma,Peeters2016}, with periodic
boundary conditions in the axial and azimuthal directions. 
In the radial direction, a sixth-order compact-finite-difference
scheme is used. Like others, I do not simulate the flow
around the entire cylinder but use a domain with reduced
size in the azimuthal direction. Previously, it has been
verified that changing the domain size has little effect
on the computed torque \citep{Brauckmann,Brauckmann2016}.
\begin{table}
\caption{Domain size in the DNSs of TCF. $L_\theta$ is the azimuthal domain size
at the centreline and $L_z$ is the axial domain size.}
\begin{center}
\begin{tabular}{lrrrrrrrr}
$\Rey$ &   $~~~400$ & $~~~1152$ & $~~~2593$ & $~~~3889$ & $~~~8750$ & $~~~19\,688$ & $~~~29\,531$ & $~~~40\,000$ \\
$L_\theta/d$ & $6\pi$ & $6\pi$ & $6\pi$ & $6\pi$ & $\pi$ & $\pi$ & $\pi$ & $3\pi$ \\   
$L_z/d$      & $2$ & $2$ & $2$ & $2$ & $2$ & $2$ & $2$ & $4\pi/3$ \\
\end{tabular}
\end{center}
\label{sim_par}
\end{table}
The computational domain size in the DNSs of TCF, listed in Table \ref{sim_par},
is basically the same as in the DNSs of \cite{Brauckmann} up to $\Rey=29\,531$
and wide enough to capture at least one pair of counter-rotating
Taylor vortices. In the DNSs at $\Rey=40\,000$ the domain is significantly larger.

The resolution increases with $\Rey$ to keep
the grid spacing in terms of viscous wall units within 
acceptable bounds. At $\Rey=40\,000$
the streamwise and spanwise
grid spacing in the PCF DNSs is
$\Delta x^+ \leq 13$ and $\Delta z^+ \leq 6.5$, respectively, and in the TCF DNSs
the azimuthal and axial grid spacing is 
$\Delta x^+ =12.4$ and $\Delta z^+ =5.9$, respectively, at $\Ro=0.3$ and smaller at higher $\Ro$.
This is the grid spacing in terms of Fourier modes and viscous wall
units, comparable to the grid spacing in other well-resolved
DNSs of wall flows \citep{Lee}.
The number of Chebyshev modes or radial grid points 
with near-wall clustering is 192 or more at this $\Rey$.

The DNSs of PCF and TCF
are either initialized with perturbations to trigger vortices or
with the fields of a DNS at another $\Ro$. They are
run for a sufficiently long time to reach a statistically stationary
state and then run for a long period to obtain well-converged statistics. 
I have verified that proceeding the DNSs does not 
change the computed Nusselt numbers.
For several DNSs of PCF I have also validated that changing the domain size and
resolution does not affect the results. The DNS results of TCF for
$\Nu_m$ agree well with previous DNSs of \cite{Brauckmann} and \cite{Ostilla}.
These validations are presented in the supplementary material.

\section{Results}
Figure \ref{nus}({\it a},{\it b}) shows that momentum transfer in terms of $\Nu_m$ naturally grows with $\Rey$ 
but also varies with $\Ro$ in the DNSs owing to changing flow features \citep{Salewski,Grossmann}.
\begin{figure}
\begin{center}
\includegraphics[width=13.5cm]{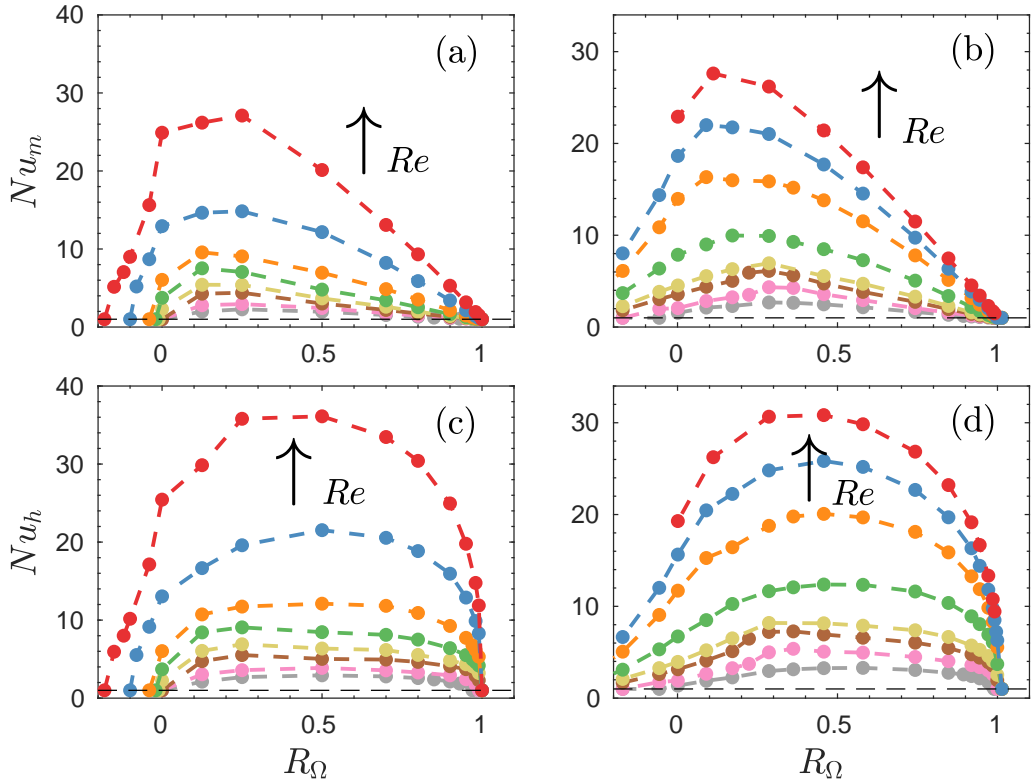}
\end{center}
\caption{({\it a}) $\Nu_m$ and ({\it c}) $\Nu_h$ for PCF.
({\it b}) $\Nu_m$ and ({\it d}) $\Nu_h$ for TCF. Each line/colour represents a different constant $\Rey$
and dots denote DNS results. 
The horizontal dashed line marks $\Nu_m,Nu_h =1$. 
Arrows show trends for increasing $\Rey$
listed in section \ref{num}.\label{nus}}
\end{figure}
At $\Ro=0$ PCF is linearly stable yet turbulent when $\Rey \gtrsim 1600$ 
owing to subcritical transition and therefore $\Nu_m>1$. 
TCF is linearly unstable and $\Nu_m >1$ 
if $\Ro=0$ at all $\Rey$ \citep{Dubrulle}. 
In both flows $\Nu_m$ first grows with $\Ro$ due to destabilization
by anticyclonic rotation and then declines towards unity 
for $\Ro\rightarrow 1$ when the flow approaches the linearly stability limit $\Roc$
and relaminarizes \citep{Dubrulle}. Disturbances and turbulence 
cannot sustain beyond $\Roc$, 
even at higher $\Rey$ \citep{Ostilla2014}.
Momentum transport is maximal around
$\Ro=0.2$ in PCF and around $\Ro=0.3$ to 0.1 at low to high $\Rey$ in TCF, 
consistent with previous numerical \citep{Salewski,Brauckmann,Brauckmann2016}
and experimental observations \citep{Gils}. This broad maximum is linked
to intermittent bursts in the outer layer in TCF and
to strong vortical motions in PCF \citep{Brauckmann,Brauckmann2016}. 
Another narrow maximum in $\Nu_m$ caused by shear instabilities
appears in PCF at $\Ro \approx 0.02$ when 
$\Rey \gtrsim 10^4$ \citep{Brauckmann}, but my DNSs 
do not cover this narrow region near $\Ro=0$ and therefore do not reveal this
second maximum. With rising $\Rey$ the narrow overtakes the
broad maximum that disappears in TCF with $\eta = 0.91$ if $\Rey$ is higher
than in my DNSs \citep{Ezeta}.

\begin{figure}
\begin{center}
\includegraphics[height=4.8cm]{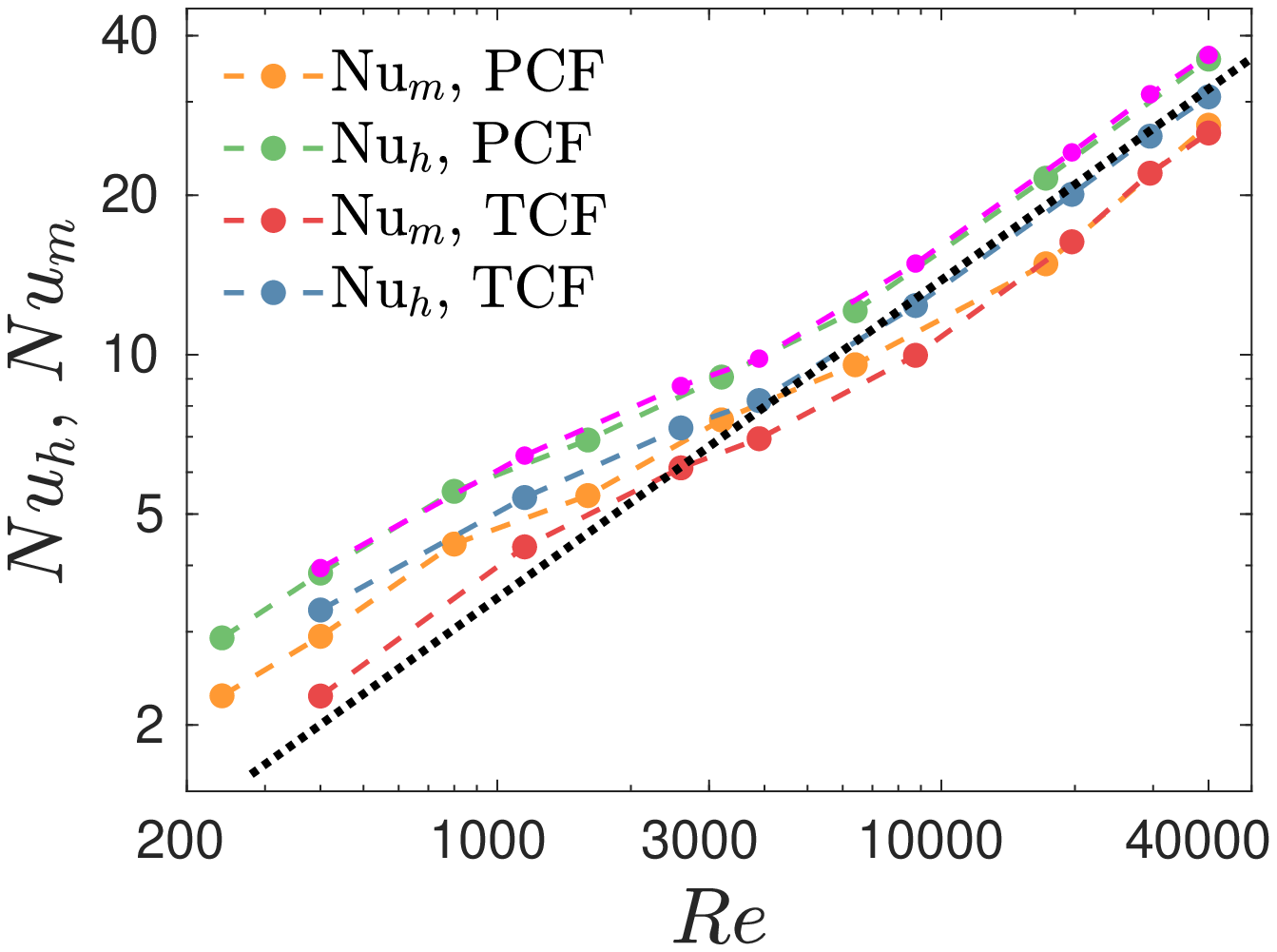}
\end{center}
\caption{Maximum value of $\Nu_m$ and $\Nu_h$ over
$\Ro$ at fixed $\Rey$
in PCF and TCF as function of $\Rey$.
The dotted black line shows the scaling 
$\Nu\sim\Rey^{0.6}$.\label{scaling}}
\end{figure}

Heat transfer in terms of $\Nu_h$ behaves similarly as $\Nu_m$ at low $\Ro$ but differently  
at higher $\Ro$ (figure \ref{nus}{\it c},{\it d}). Its maximum is higher and
at higher $\Ro$ for almost all $\Rey$,
demonstrating that flow structures causing optimal momentum transport
do not necessarily cause optimal heat transport.
At higher $\Rey$, $\Nu_h$ is maximal near $\Ro=0.5$ in both PCF and TCF
and then sharply declines when $\Ro\rightarrow 1$ and the flow relaminarizes. 
This means that 
in a laboratory instead of rotating frame of reference,
maximal momentum transfer in higher $\Rey$ TCF 
happens with moderately counter-rotating whereas
maximal heat transfer happens with co-rotating 
inner and outer cylinders.
The growth of the maximum $\Nu_m$ and $\Nu_h$ 
over all $\Ro$ at fixed $\Rey$
show similar trends with $\Rey$
in PCF and TCF and follows $\Nu_m,\Nu_h \sim Re^{0.6}$ at higher
$\Rey$ (figure \ref{scaling}).
Experiments show for $\Rey > 3.10^4$ the scaling 
$\Nu_m \sim \Rey^{0.78}$ for all $\Ro$ in TCF \citep{Ostilla2014,Ostilla2014b}.
It is therefore possible that at higher $\Rey$ the maximum $\Nu_h$ follows a
different scaling than $\Nu_h \sim \Rey^{0.6}$ observed here.

\begin{figure}
\begin{center}
\includegraphics[width=13.5cm]{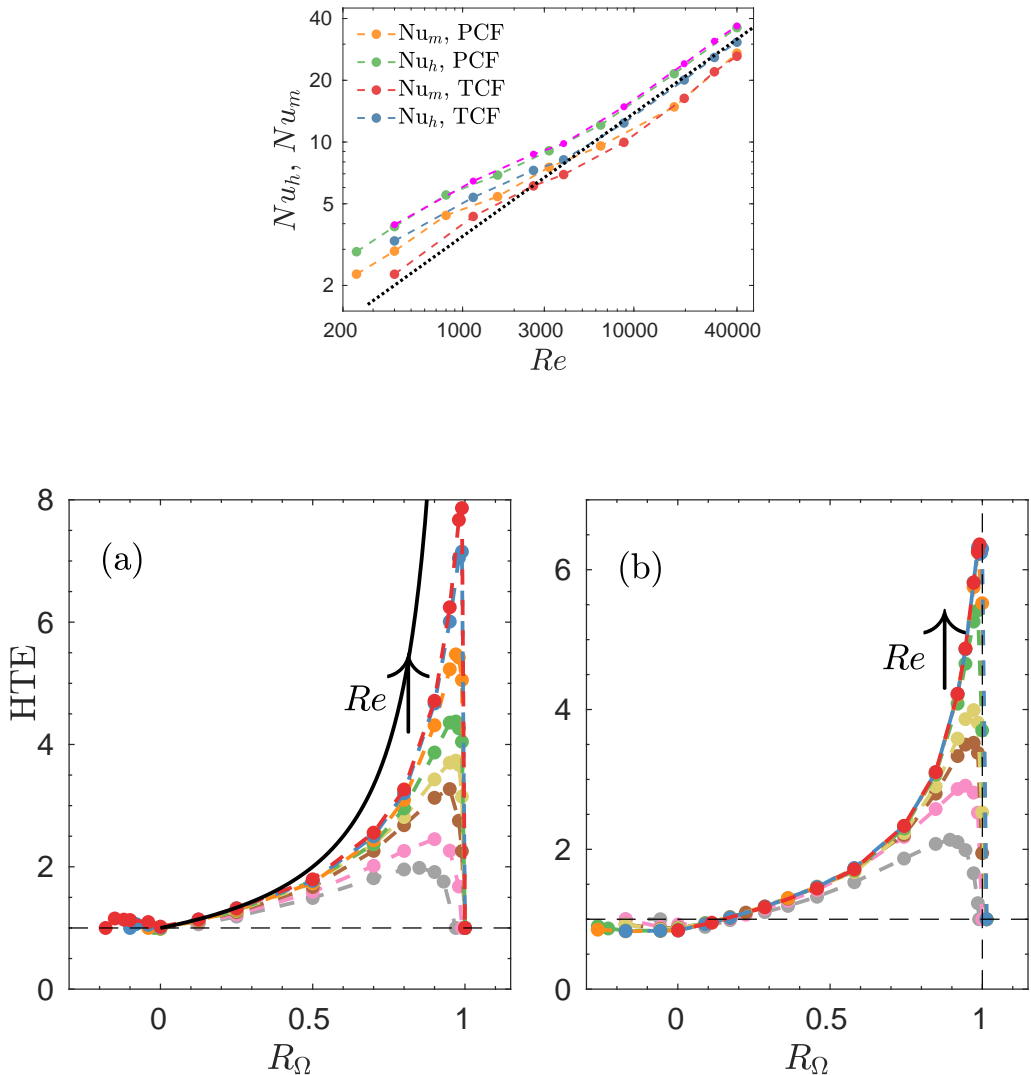}
\end{center}
\caption{HTE as function of $\Ro$ in ({\it a}) PCF and ({\it b}) TCF.
Each line/colour represents a different constant $\Rey$.
Arrows show trends for increasing $\Rey$ listed in section \ref{num}
and dots signify DNS results. The black line in ({\it a})
shows the high-$\Rey$ limit.\label{hte}}
\end{figure}

\begin{figure}
\begin{center}
\includegraphics[height=4.7cm]{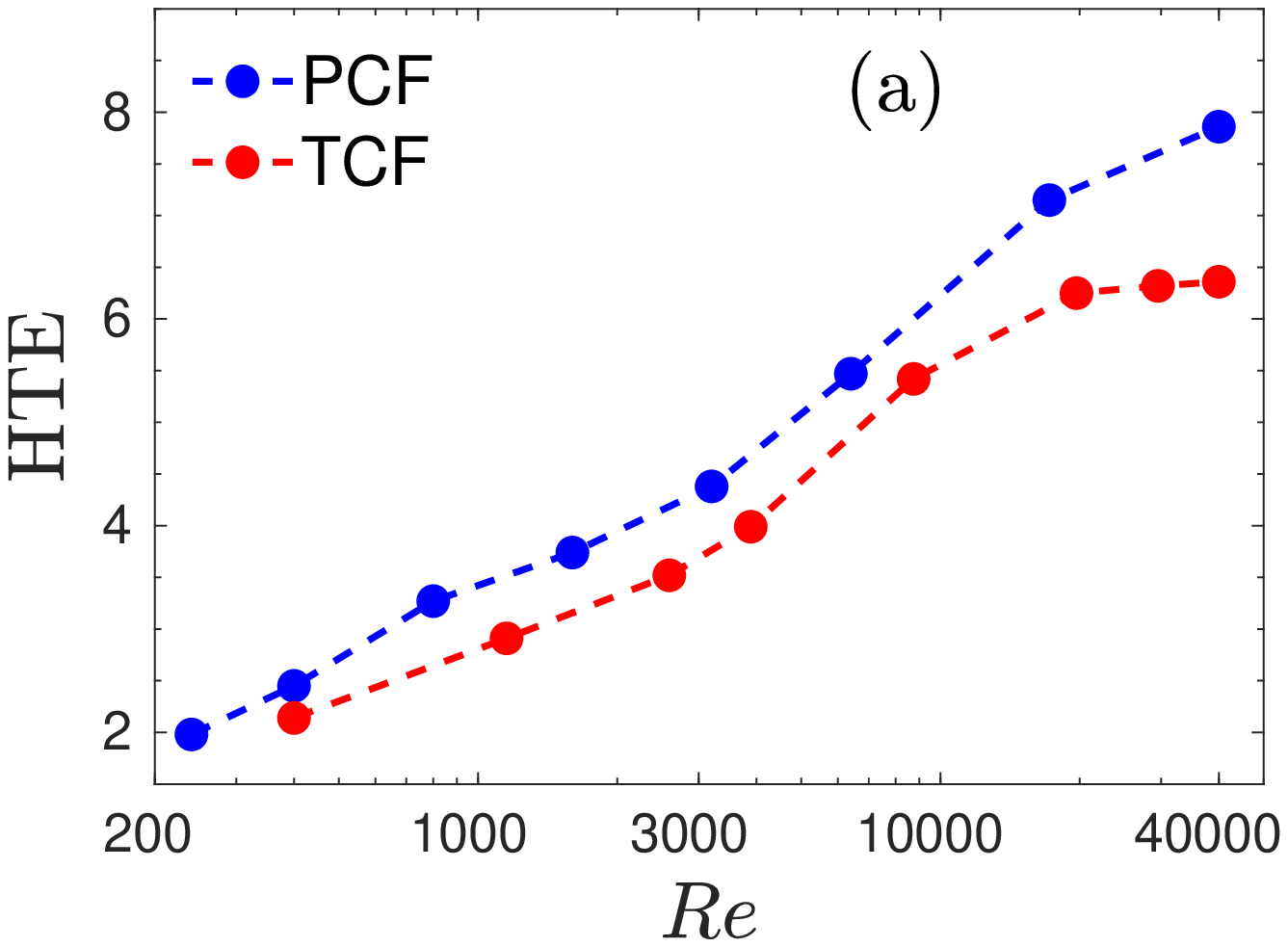}
\hskip1mm
\includegraphics[height=4.7cm]{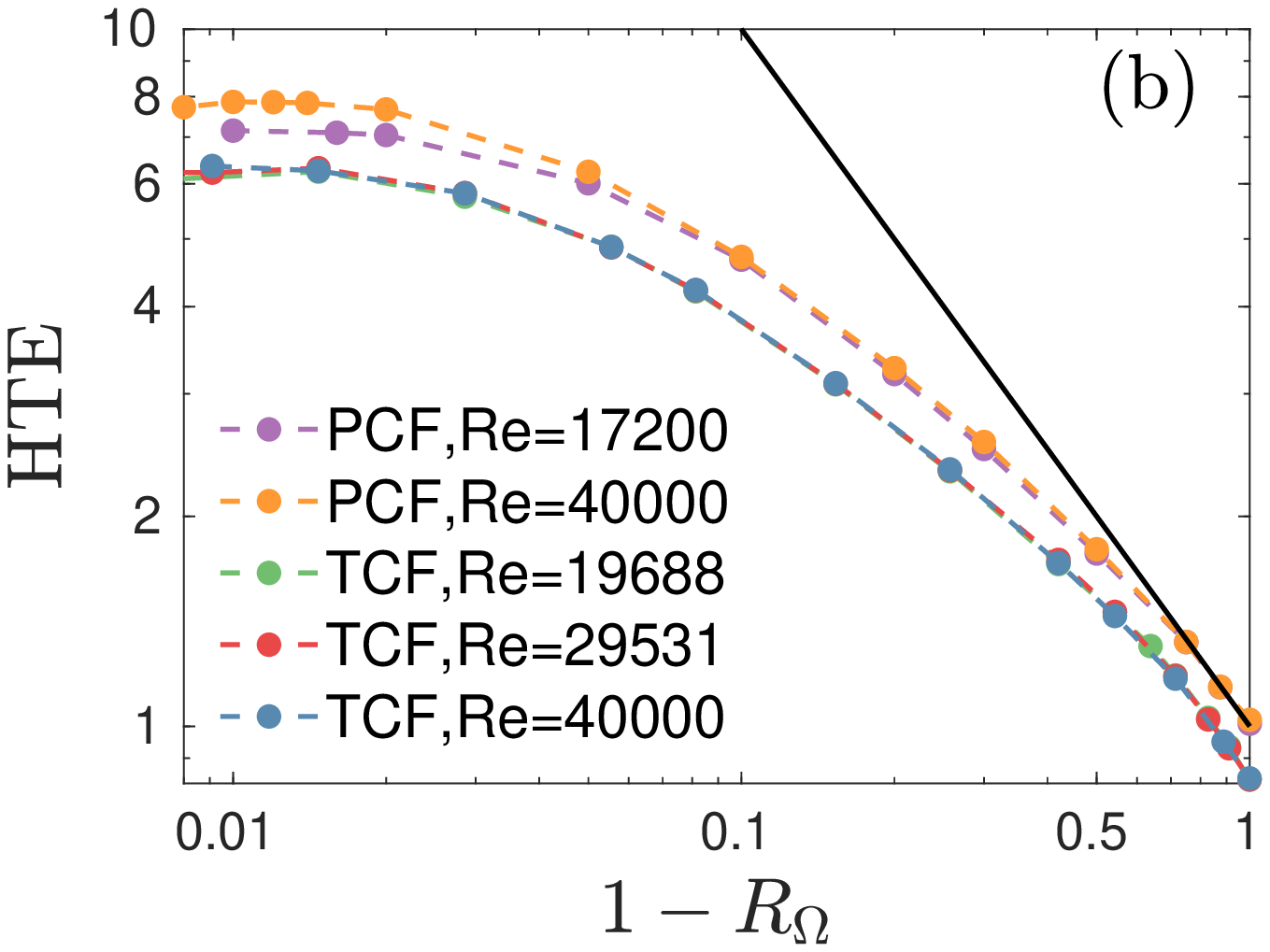}

\vskip4mm
\includegraphics[height=5.2cm]{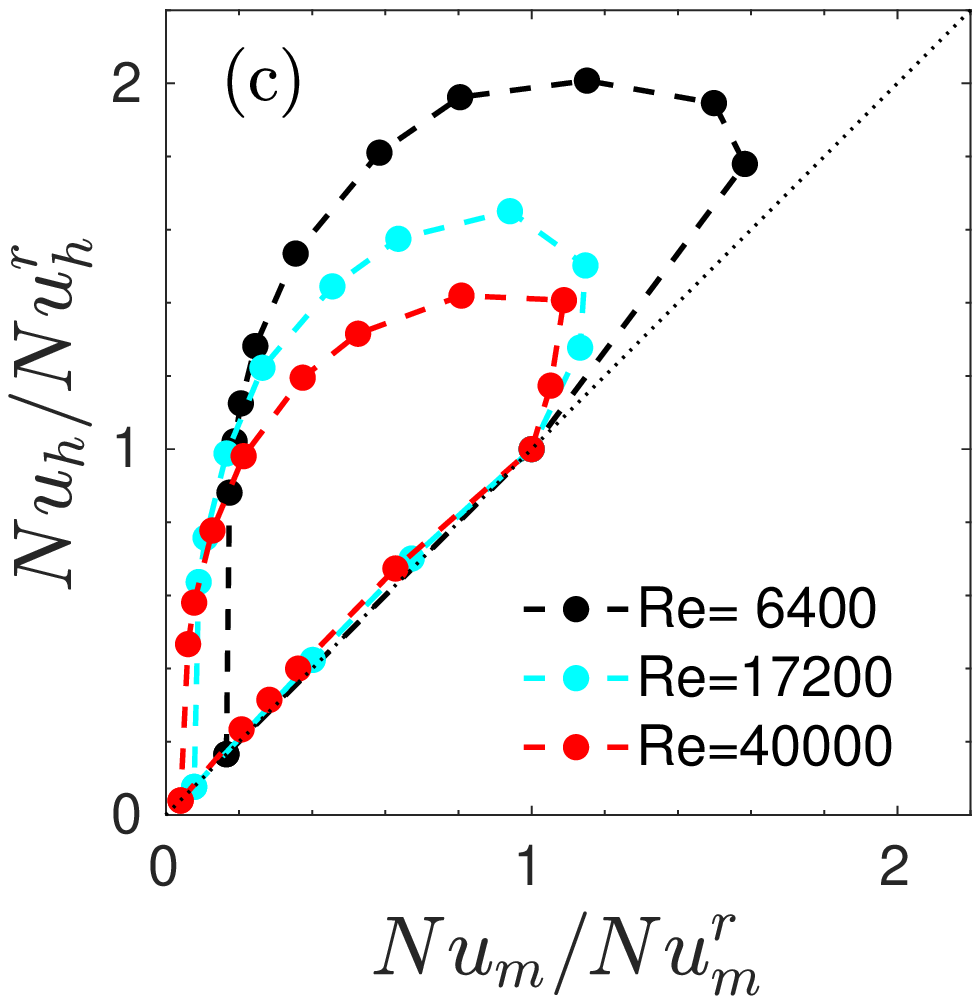}
\hskip14mm
\includegraphics[height=5.2cm]{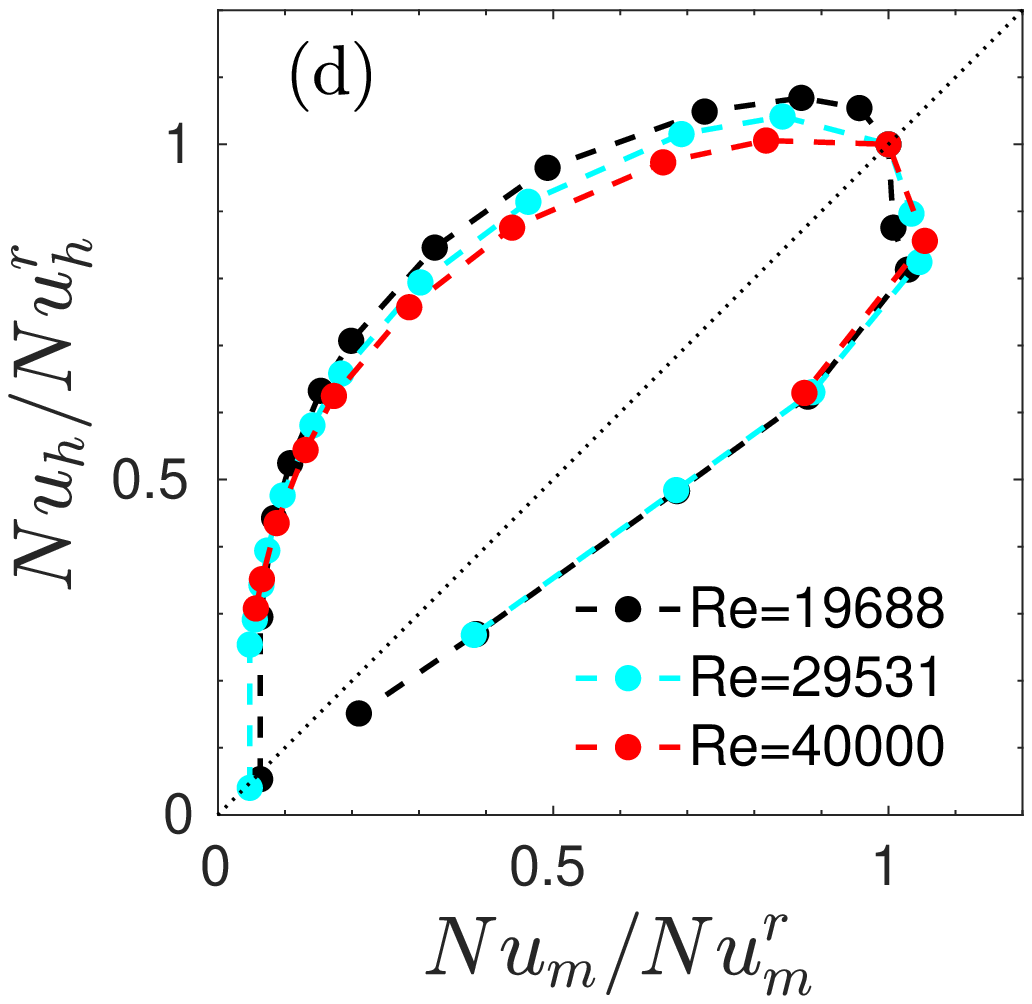}
\end{center}
\caption{({\it a}) Maximum $\mbox{HTE}$ over
$\Ro$ at fixed $\Rey$
in PCF and TCF as function of $\Rey$.
({\it b}) Log-log plot of $\mbox{HTE}$ versus $1-\Ro$ in PCF
at $\Rey = 17\,200$ and $40\,000$
and TCF at $\Rey=19\,688$, $25\,531$ and $40\,000$. 
The black solid line shows the high-$\Rey$ limit $\mbox{HTE}=1/(1-\Ro)$.
$\Nu_h/\Nu^r_h$ versus $\Nu_m/\Nu^r_m$ at fixed $\Rey$ for 
({\it c}) PCF and ({\it d}) TCF.
$\Nu^r_{m,h}$ is the reference Nusselt number corresponding
to the non-rotating case ($\Ro=0$) in PCF and the fixed outer
cylinder case ($\Ro=1-\eta=0.29$) in TCF.
$\Ro$ increases in the counterclockwise direction.
\label{hte_log}}
\end{figure}

The ratio $\mbox{HTE}=\Nu_h/\Nu_m$,
shown in figure \ref{hte}, is a measure of heat transfer efficiency
since $\Nu_m$ is proportional to the power input \citep{Gils}.
%
An equivalent measure is considered in heat transfer optimization studies
by \cite{Yamamoto,Motoki,Kaithakkal}.
A high similarity between momentum and heat transport can be expected at 
$\Ro=0$ in PCF because $\Prandtl=1$ and momentum 
and heat transport are similarly forced. 
This is vindicated by the DNSs; the difference between
$\Nu_h$ and $\Nu_m$ is not more than 2\% at all $\Rey$ and accordingly $\mbox{HTE}\simeq 1$,
meaning that the Reynolds analogy perfectly applies. 
$\mbox{HTE}$ is somewhat smaller in TCF at $\Ro=0$ because $\Prandtl < 1$
but still near unity so that the Reynolds analogy practically holds.  
Clear differences in heat and momentum transport emerge for increasing $\Ro$.
HTE rapidly grows with $\Ro$ in PCF and TCF
and reaches a maximum around $\Ro\approx 0.85 - 0.99$ at low to high $\Rey$
before abruptly dropping to unity for $\Ro \geq 1$.
Its maximum grows from about two at the lowest $\Rey$
to eight and more than six at $\Rey=40\,000$ in PCF and TCF, respectively
(figure \ref{hte} and figure \ref{hte_log}{\it a}).
In TCF the maximum HTE seems to level off at higher $\Rey$ while in PCF it still grows.
TCF with a fixed outer and rotating inner cylinder in a laboratory
frame corresponds to $\Ro=1-\eta=0.29$ and has a $\mbox{HTE} \leq 1.18$ 
over all $\Rey$ considered here, much less than the maximum possible $\mbox{HTE}$.
Note that Couette flow is linearly unstable very near $\Ro=1$ \citep{Nagata1990,Esser},
so flow motions can be sustained even very near $\Ro=1$.
Figure \ref{hte_log}({\it b}) shows that the $\mbox{HTE}$ has a similar trend in PCF and TCF
near $\Ro=1$ for $\Rey \geq 17\,200$.
DNS results for TCF at the three $\Rey$ are hardly
distinguishable since they overlap, although the
computational domain sizes are different, suggesting the results
are indifferent to this aspect.
Some of the dissimilarities between heat transfer in PCF and TCF can be attributed
to the different $Pr$ in the two cases. DNSs of heat transfer in turbulent
channel flow support that $\Nu_h \sim Pr^{1/2}$ for $Pr$ near unity
\citep{Pirozzoli}, indicating that $\Nu_h$ and $\mbox{HTE}$ are a factor
$0.7^{-1/2}$ larger in TCF if $Pr=1$, as in PCF,
instead of $0.7$, as used here.
The differences observed in e.g. figures \ref{scaling} and \ref{hte_log} 
between PCF and TCF results indeed largely disappear
(not shown here) and $\Nu_h$ is almost unity in TCF at $\Ro=0$ after this scaling.

Results 
for $\mbox{HTE}$ can be compared to theoretical optimal 
heat transport for $Pr=1$ in PCF with an arbitrary body force
calculated by \cite{Motoki}, who maximized heat and
momentum transport dissimilarity by optimizing the ratio
of total scalar dissipation to total energy dissipation $\varepsilon_S/\varepsilon_E$.
For rotating PCF, with no energy input by the Coriolis force,
$\mbox{HTE}$ is equivalent to $\varepsilon_S/\varepsilon_E$.
If $\Rey \leq 800$ and rotating PCF is streamwise-invariant as discussed later,
$\varepsilon_S/\varepsilon_E$ calculated by \cite{Motoki}
is nearly equal to the maximal $\mbox{HTE}$
found here, implying that heat transport
in rotating PCF is near optimal. At higher $\Rey$ results diverge;
the optimal $\varepsilon_S/\varepsilon_E$ calculated by \cite{Motoki}
at $\Rey=1600$, $6400$ and $40\,000$, shown in their figure 7, is respectively
1.6, 2.6 and 3.1 times higher than the maximal $\mbox{HTE}$ in rotating PCF.
Their optimal heat transfer rate is also higher than in turbulent PCF
without body forces.
Thus, even though the dissimilarity between heat and momentum can be large
in rotating PCF, it is theoretically possible to enhance it further
through other body forces. Another factor, which can contribute to
the observed difference, is that rotating
PCF is unsteady and turbulent for $\Rey > 800$ whereas
\cite{Motoki} only considered optimal heat transport in steady PCF.

$\mbox{HTE}$ defines the heat to momentum transfer ratio.
In applications of advective heat/mass transfer one may 
seek to optimize other variables owing to constrains \citep{Webb,Hesselgreaves,Yamamoto},
for example, to enhance heat transfer for equal power 
input or to reduce power input for equal heat transfer.
Figure \ref{hte_log}({\it c}) and ({\it d}) therefore show $\Nu_h/\Nu^r_h$ versus 
$\Nu_m/\Nu^r_m$ for PCF and TCF at fixed $\Rey$ and varying $\Ro$
for the highest $\Rey$.
Here, $\Nu^r_{m,h}$ are Nusselt numbers in the reference case at the same $\Rey$.
For PCF the non-rotating case ($\Ro=0$) is taken as reference whereas
for TCF the fixed outer cylinder case in a laboratory frame
($\Ro=1-\eta=0.29$) is taken as reference
because of its experimental relevance.
Cases left of the dotted line have a higher $\mbox{HTE}$ than the reference case.
Some $\Ro$ cases in PCF have the same wall-to-wall 
momentum transfer ($\Nu_m/\Nu^r_m \simeq 1$) 
but much higher heat transfer per unit area ($\Nu_h/\Nu^r_h > 1$), or
the same heat transfer ($\Nu_h/\Nu^r_h \simeq 1$) but much
lower momentum transfer ($\Nu_m/\Nu^r_m < 1$) than at $\Ro=0$
(figure \ref{hte_log}{\it c}).
At $\Rey = 40\,000$ and $\Ro=0.5$,
$\Nu_h/\Nu^r_h =1.4$ and $\Nu_m/\Nu^r_m =0.8$ giving a $\mbox{HTE}=1.8$, 
and at $\Ro=0.9$, $\Nu_h/\Nu^r_h \simeq 1$ and $\Nu_m/\Nu^r_m \simeq 0.21$
giving a $\mbox{HTE}=4.7$.
In TCF only a few cases have a somewhat higher heat transfer and/or lower
momentum transfer than the reference case (figure \ref{hte_log}{\it d})
since $\Nu_h$ 
is already high at $\Ro=0.29$ (figure \ref{nus}{\it d}).
However, $\mbox{HTE}$ can be much higher than 1.18 as
in the reference case, as shown before.
When $\eta$ increases we can expect
$\Nu_h/\Nu^r_h$ versus $\Nu_m/\Nu^r_m$ curves for TCF 
to resemble curves for PCF if the fixed outer cylinder case
is taken as reference since this case corresponds to $\Ro=1-\eta$,
which approaches zero.

\begin{figure}
\begin{center}
\includegraphics[height=5.2cm]{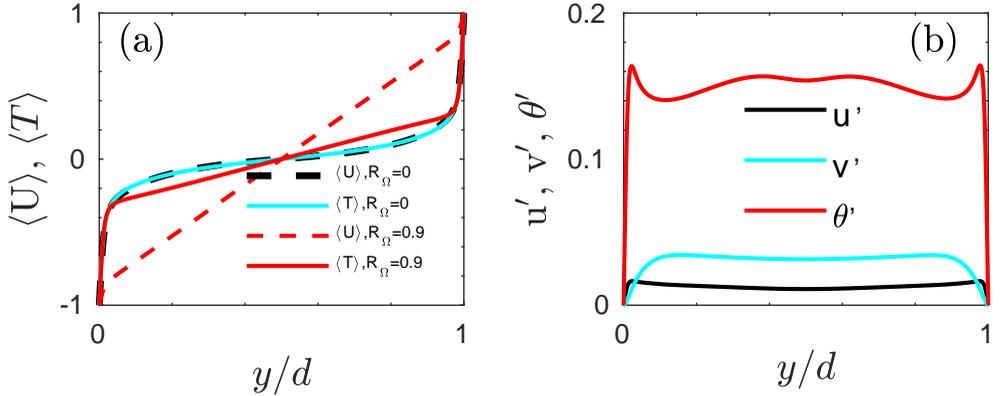}
\end{center}
\caption{({\it a}) Profiles of $\langle \mbox{U}\rangle$
and $\langle T\rangle$ in PCF at $\Rey=40\,000$ and $\Ro=0$ and 0.9.
({\it b}) Profiles of the root-mean-square of the streamwise and wall-normal
velocity fluctuations $u'$ and $v'$,
respectively, and temperature fluctuation $\theta'$ in PCF
at $(\Rey;\Ro)=(40\,000;0.98)$. Mean and fluctuating
velocity and temperature are scaled by $U_w$ and $T_w$, respectively, and $y$ is the distance to the wall
($y/d=0$ or 1 at the walls). \label{prof}}
\end{figure}

In anticyclonic rotating Couette flows heat is thus transported much 
faster than momentum, in violation of the Reynolds analogy, when approaching
the linear stability limit $\Roc\simeq 1$.
Mean velocity and temperature profiles reflect the transport anomaly;
at $\Ro=0$ these are barely distinguishable in PCF, but
when $\Ro\rightarrow 1$ the mean temperature $\langle T \rangle$ 
has a thin boundary layer and nearly linear centre profile and clearly differs
from the mean streamwise velocity $\langle \mbox{U}\rangle$, which approaches the linear
laminar profile (figure \ref{prof}{\it a}).
Here, $\langle \cdot \rangle$ denotes averaging over time and wall-parallel planes.
Mean velocity and temperature profiles in TCF are not shown but behave similarly. 
The thermal boundary layer becomes thinner with $\Rey$ and changes more rapidly 
than the velocity boundary layer for $\Ro$ near $\Roc$,
leading to a growth of the maximum HTE.

Insight into the transport anomaly and small streamwise velocity fluctuations 
and related weak momentum transport at high $\Ro$ 
is obtained by studying the action of the Coriolis force.
Consider the mean shear and Coriolis force term 
in the governing equation for 
$u$ in PCF, that is $v(2\Omega - \partial_y \langle \mbox{U} \rangle)$,
where $u$ and $v$ is the streamwise and wall-normal velocity fluctuation, respectively;
if $\Omega > 0$ the Coriolis force reduces production of $u$ by mean shear when $v\neq0$.
Note that the absolute mean vorticity $\partial_y \langle \mbox{U} \rangle - 2 \Omega \approx 0$
about the channel centre at sufficiently high $\Ro$ \citep{Brauckmann2016,Kawata}
and in the whole channel if $\Ro\rightarrow 1$. The Coriolis term in
the Reynolds stress transport equation of $\langle uu \rangle$ counterbalances then the production
term and the only term producing $\langle uu \rangle$
is the pressure-strain correlation \citep{Geert2017}.
If a fluid particle is displaced in the wall-normal direction by vortical motions
the Coriolis force basically accelerates or decelerates the particle so that its 
streamwise velocity approaches the local mean velocity.
Figure \ref{prof}({\it b}) confirms that 
$u'=\langle uu \rangle^{1/2}$ 
is small in PCF if $\Ro\rightarrow 1$ while 
$v'=\langle vv \rangle^{1/2}$ 
is larger because vortical motions survive as long as $\Ro < \Roc$
and produce a high heat flux and intense temperature fluctuations
that are not directly affected by the Coriolis force.
Observations in TCF (not shown) are again similar; the specific
angular momentum is nearly constant if $\Ro \rightarrow 1$, implying
neutral stability according to Rayleigh's criterion \citep{Brauckmann2016}
and strongly reduced azimuthal velocity fluctuations.

\begin{figure}
\begin{center}
\includegraphics[width=5.8cm]{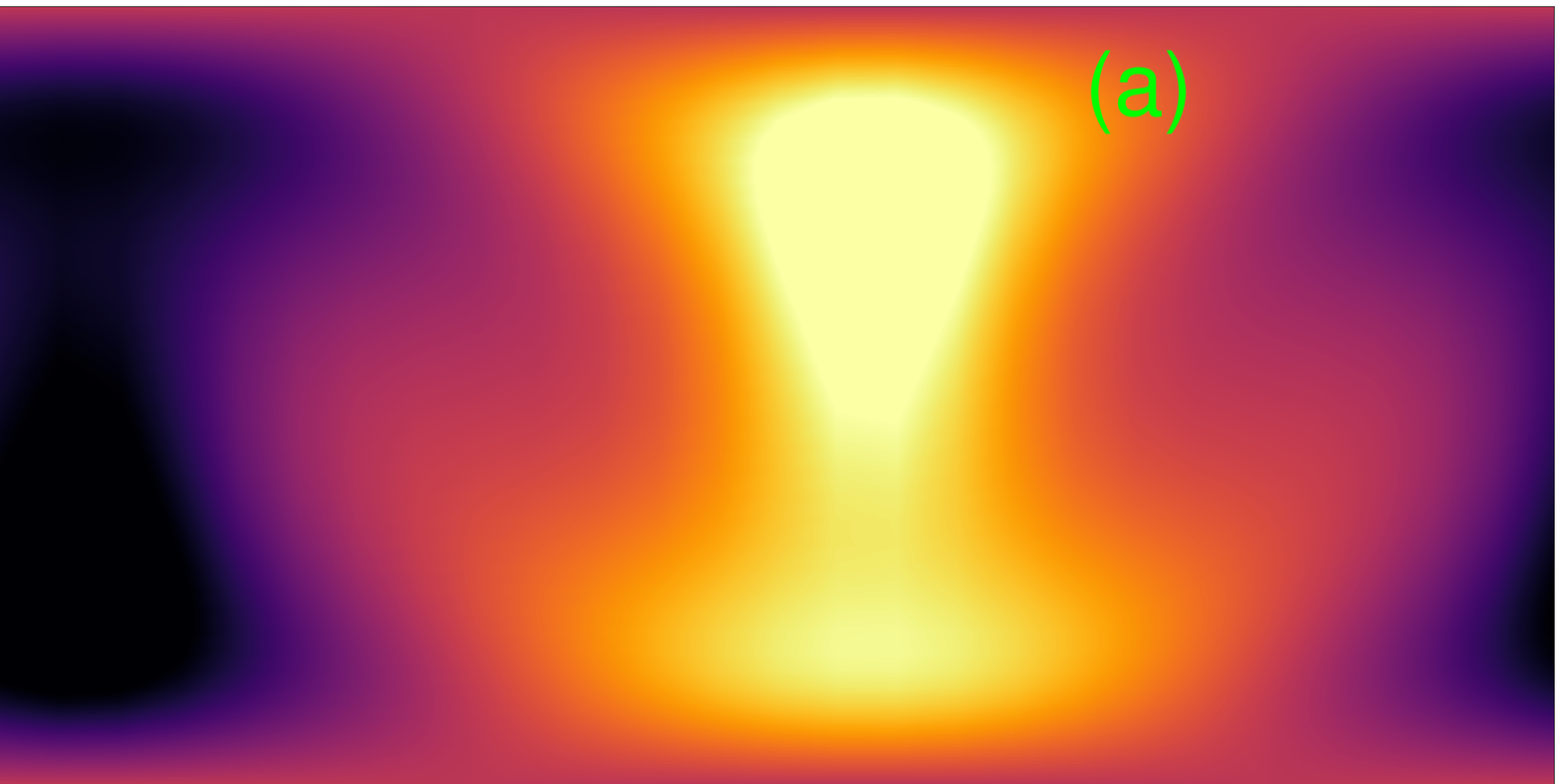}
\hskip3mm
\includegraphics[width=5.8cm]{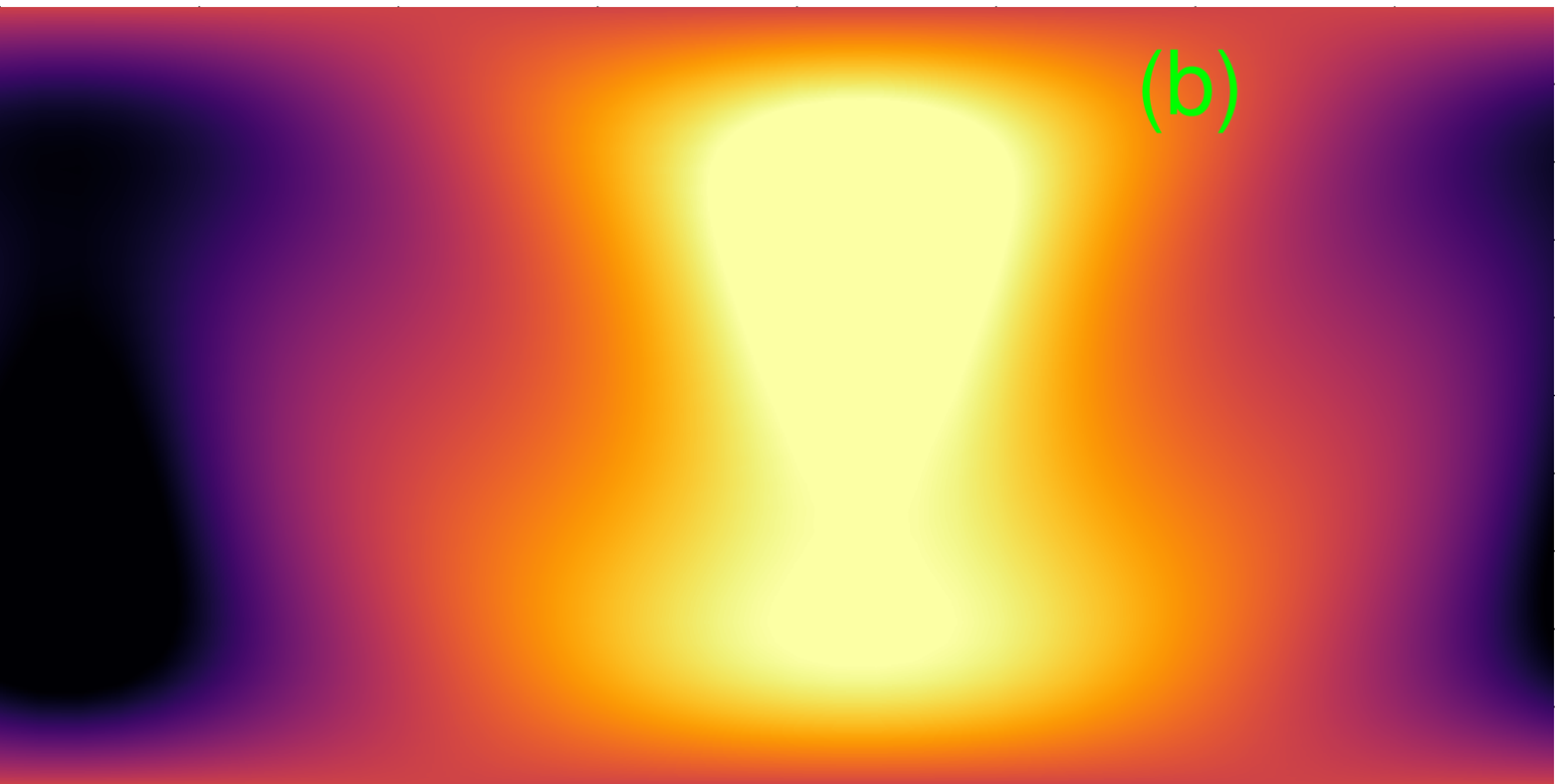}

\vskip2mm
\includegraphics[width=5.8cm]{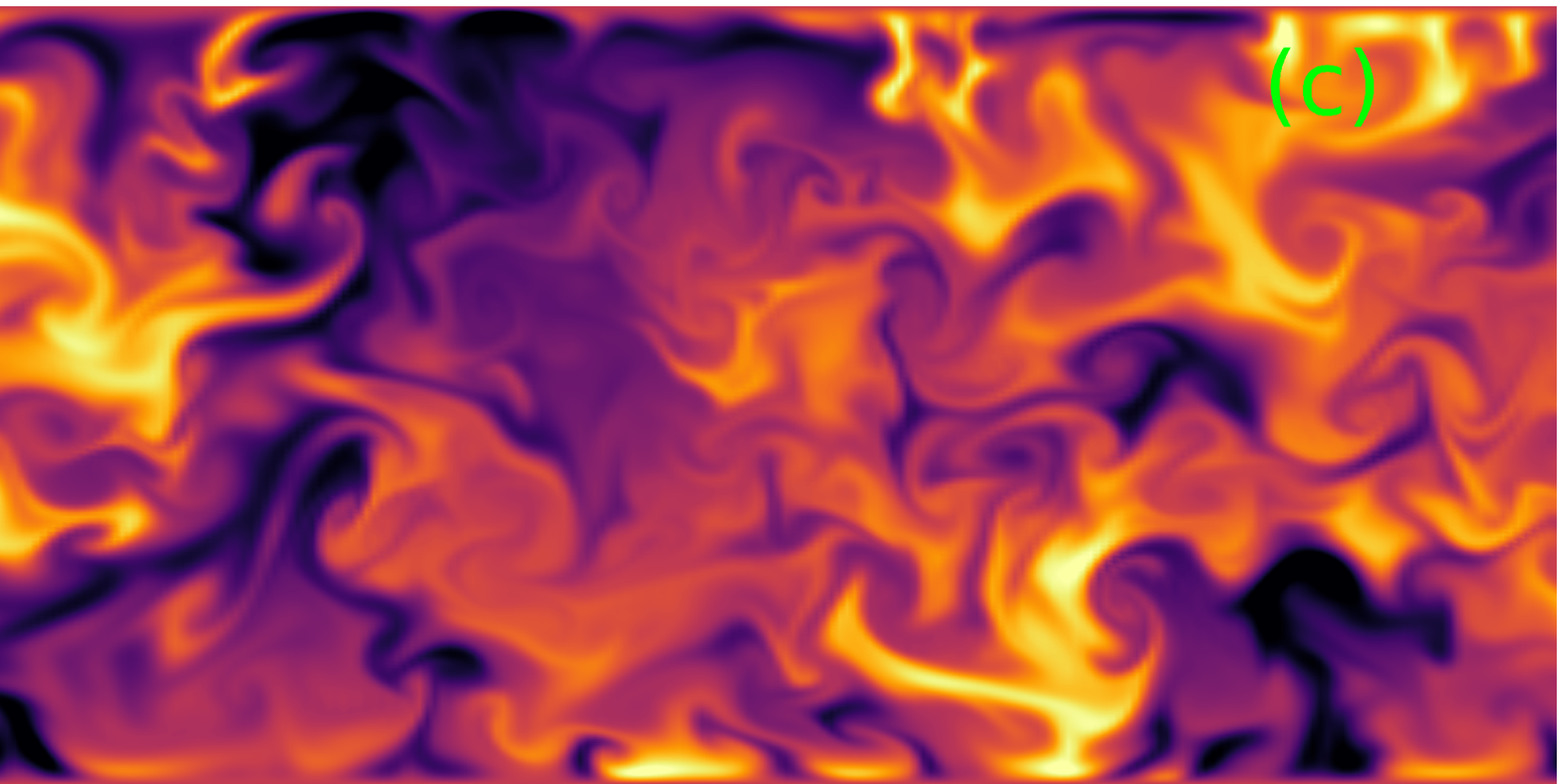}
\hskip3mm
\includegraphics[width=5.8cm]{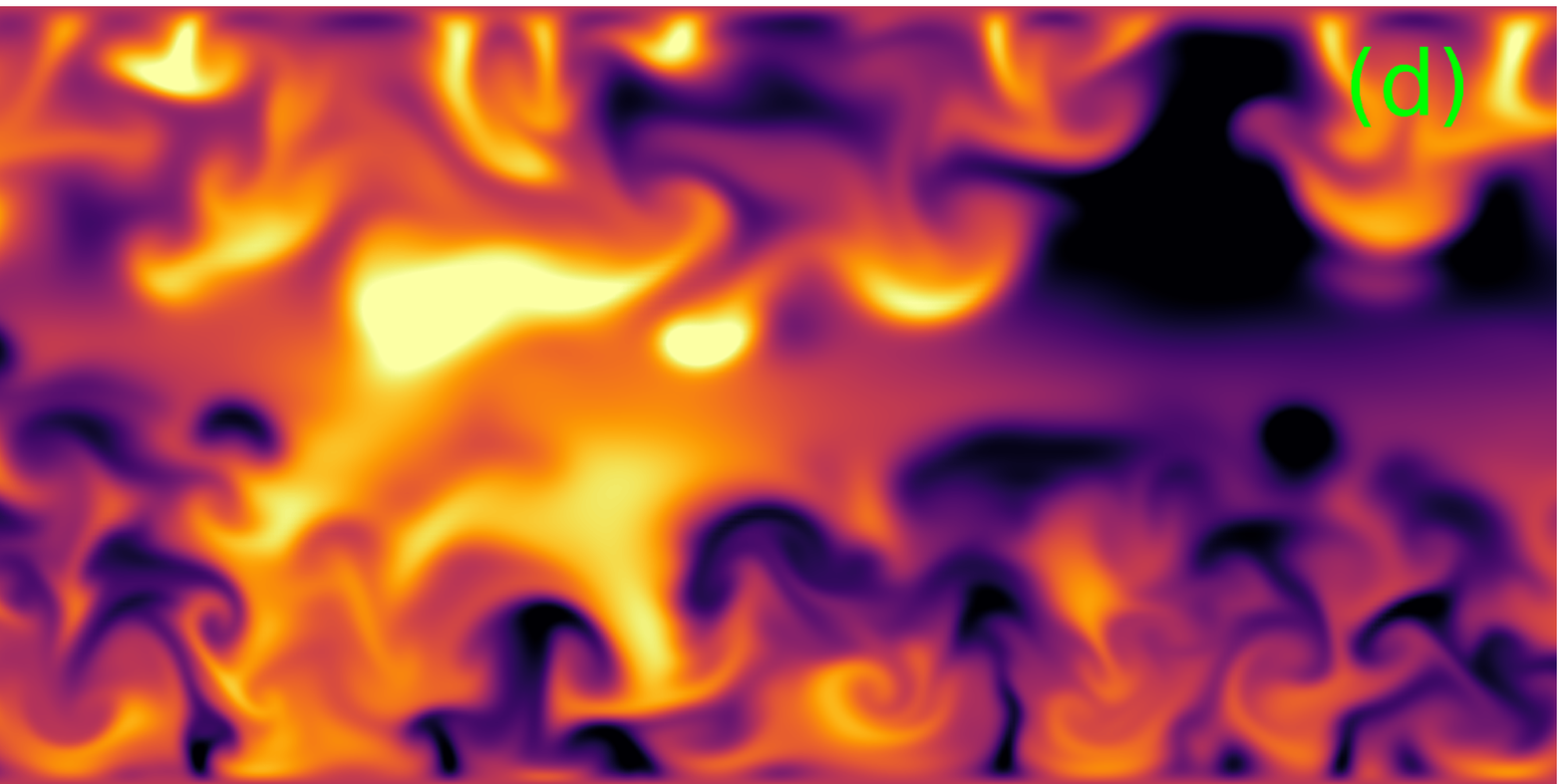}
\end{center}
\caption{Snapshot of the instantaneous temperature 
fluctuation field at ({\it a},{\it b}) $\Rey=400$
and $\Ro=0.9$ and ({\it c},{\it d}) $\Rey=40\,000$ and $\Ro=0.99$ in ({\it a},{\it c}) PCF
and ({\it b},{\it d}) TCF in a cross-stream plane. In ({\it b}) the full
and ({\it a},{\it c},{\it d}) a part of the
spanwise/axial domain is shown.\label{vis}}
\end{figure}

Steady streamwise-invariant Taylor vortices are present at low $\Rey$.
These are seen in visualizations of the flow field
(not shown here) and indicated by the visualizations of
the temperature field in figure \ref{vis}({\it a},{\it b}).
These vortices appear above the stability limit \citep{Nagata1990,Esser}
and echo structures producing optimal heat transport
in theoretical studies of PCF \citep{Motoki}.
They transport considerable heat but little momentum
since streamwise velocity fluctuations are small when $\Ro\rightarrow1$, as discussed above.
For streamwise-invariant PCF with $\Prandtl=1$ one 
can further quantify this and derive from (\ref{nse}) and (\ref{te})
\begin{equation}
\hat{u}/U_w=(1-\Ro)\hat{\theta}/T_w,
\label{fluc_rel}
\end{equation}
where $\hat{u}$ and 
$\hat{\theta}$ are the streamwise velocity and temperature deviations 
from the laminar situation, respectively. Further, using a variable 
transformation as in \cite{Zhang} gives
\begin{equation}
\mbox{HTE}=1+\frac{\Ro}{1-\Ro}\,\frac{\Nu_m-1}{\Nu_m}.
\label{hte_rel}
\end{equation}
\cite{Eckhardt2020} derive exact relations between heat and momentum
transport in two-dimensional Rayleigh-B{\'e}nard convection and rotating PCF
that are equivalent to (\ref{hte_rel}).
Relations (\ref{fluc_rel}) and (\ref{hte_rel}) 
are exact as long as $\Rey \leq 800$ and PCF is
streamwise-invariant. Equation (\ref{fluc_rel}) shows 
that $\hat{u}$ declines relative to $\hat{\theta}$ 
when $\Ro \rightarrow 1$ and (\ref{hte_rel}) shows that $\mbox{HTE}=1$ if $\Ro=0$ but
$\mbox{HTE}>1$ if $0<\Ro<1$ and $\Nu_m > 1$,
so heat is transported more efficiently than momentum.
Equation (\ref{hte_rel}) further suggests a growing $\mbox{HTE}$ with $\Rey$
as $\Nu_m$ increases. 
In the high $\Rey$-limit, $\Nu_m \rightarrow \infty$ and consequently
$\mbox{HTE} \rightarrow 1/(1-\Ro)$ for streamwise-invariant PCF.
The simulated $\mbox{HTE}$ is lower because of finite $\Rey$ 
and turbulence, 
but for $0\leq \Ro \leq 0.25$ when quasi two-dimensional
streamwise vortices dominate transport \citep{Brauckmann2016}
this high-$\Rey$ limit, shown by the black solid line in figure \ref{hte}({\it a}),
closely matches DNSs.

In TCF both curvature and rotation play a role.
In the rotating reference frame, the centripetal/Coriolis 
acceleration terms caused by streamline curvature in the non-dimensional 
equations of motion for TCF in cylindrical coordinates 
scale with the curvature number $R_C = (1-\eta)/\sqrt{\eta}$, 
see the supplementary material and \cite{Dubrulle} and \cite{Brauckmann2016}.
Curvature effects should therefore disappear if $\eta \rightarrow 1$
and consequently $R_C \rightarrow 0$. 
TCF properties are indeed similar to those of PCF 
and profiles of $Nu_m$ as function of $\Ro$ collapse 
for $\eta \geq 0.9$ if $\Rey$ is equal, see \cite{Brauckmann2016}, 
confirming that $\Rey$ and 
$\Ro$ appropriately describe Couette flows and the TCF to PCF transition.
We can then expect also heat transfer in PCF and TCF 
with $\eta \geq 0.9$ to be similar if both $\Rey$ and $\Ro$ are equal.
By contrast, when $\eta < 0.9$ and $\Ro$ is low, curvature effects are important. 
TCF is then continuously turbulent in the inner partition while 
strongly intermittent in the outer partition 
due to a stabilizing curvature influence, 
see \citep{Brauckmann2016}, and this affects momentum transfer.

However, beyond a critical $\Ro$, TCF is fully turbulent
in the outer partition as well and curvature effects are again 
less important.
Consider the ratio of the curvature and rotation terms in the 
equations of motion given by $R_C U/(2\Ro)$, 
where $U$ is the streamwise velocity scaled by $U_w$, 
see the supplementary material for details.
When $\Ro$ is sufficiently high it is less than unity and 
therefore rotation influences should dominate even if $\eta < 0.9$.
For $\Ro \gtrsim 0.2$, $Nu_m$ collapses for $\eta \geq 0.71$ and 
$\Rey$ as well as $\Ro$ are equal, as shown by 
\cite{Brauckmann2016} for moderate $\Rey$,
which agrees with that idea. Also in the present study, 
$Nu_m$ as function of $\Ro$ in PCF 
and TCF collapse for $\Rey=400$ and $40\,000$ and $\Ro \gtrsim 0.3$.
This is not explicitly shown but can be inferred by comparing
figure \ref{nus}({\it a}) and ({\it b}).
In the other cases $\Rey$ is different, which complicates a comparison.
When rotation influences dominate and $Nu_m$ is similar, we can expect
that also $Nu_h$ is similar in PCF and TCF. That appears to be 
true for the present cases once taking into account 
the difference in $Pr$. 
For lower $\Ro$ curvature effects are noticeable and cause differences
in $Nu_m$ and $Nu_h$ in TCF and PCF, see figure \ref{nus}.
$\mbox{HTE}$ seems to be less affected by curvature for $\Ro \geq 0$,
although it could possibly cause a difference in
heat and momentum transfer.

To summarize, streamline curvature has a noticeable effect
on heat and mass transfer when it partly stabilizes TCF
for $\eta < 0.9$ and sufficiently low $\Ro$. When $\eta \geq 0.9$
or when $\Ro$ is sufficiently high and stabilization of
TCF in the outer partition does not occur,
curvature effects appear to have
a small or negligible influence on heat and momentum transfer.

\begin{figure}
\begin{center}
\includegraphics[height=4.0cm]{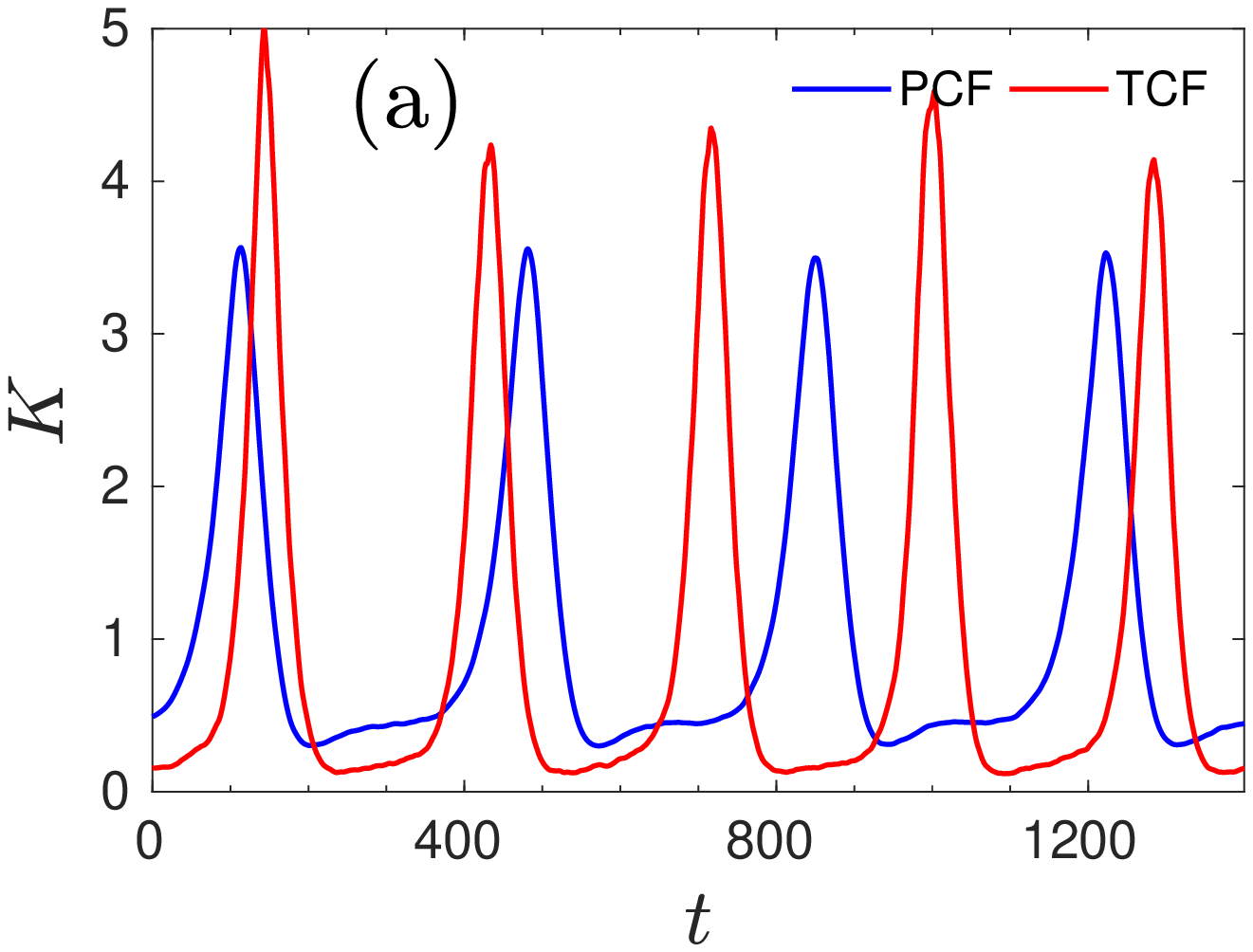}
\hskip8mm
\includegraphics[height=4.3cm]{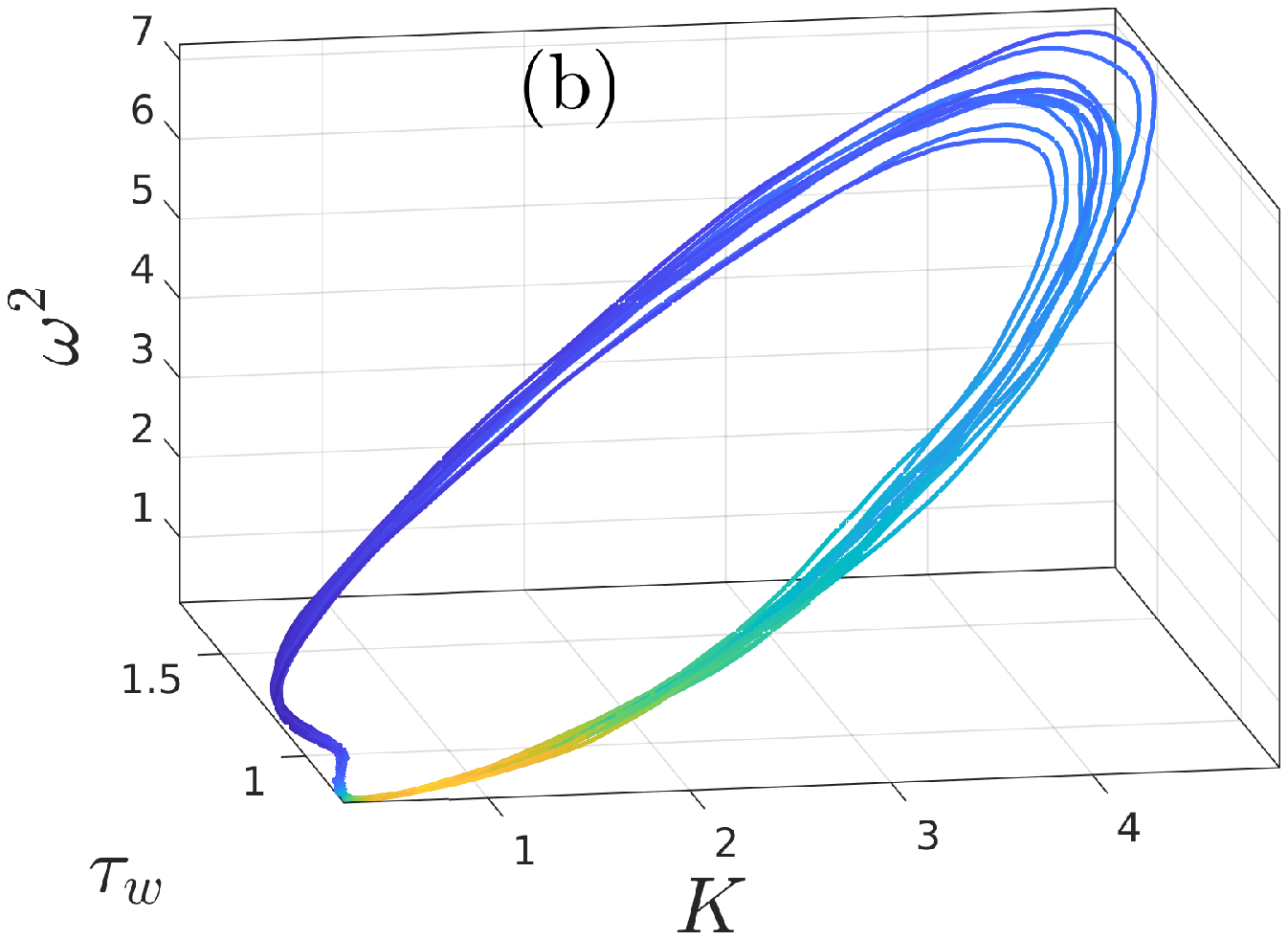}

\vskip3mm
\includegraphics[height=4.5cm]{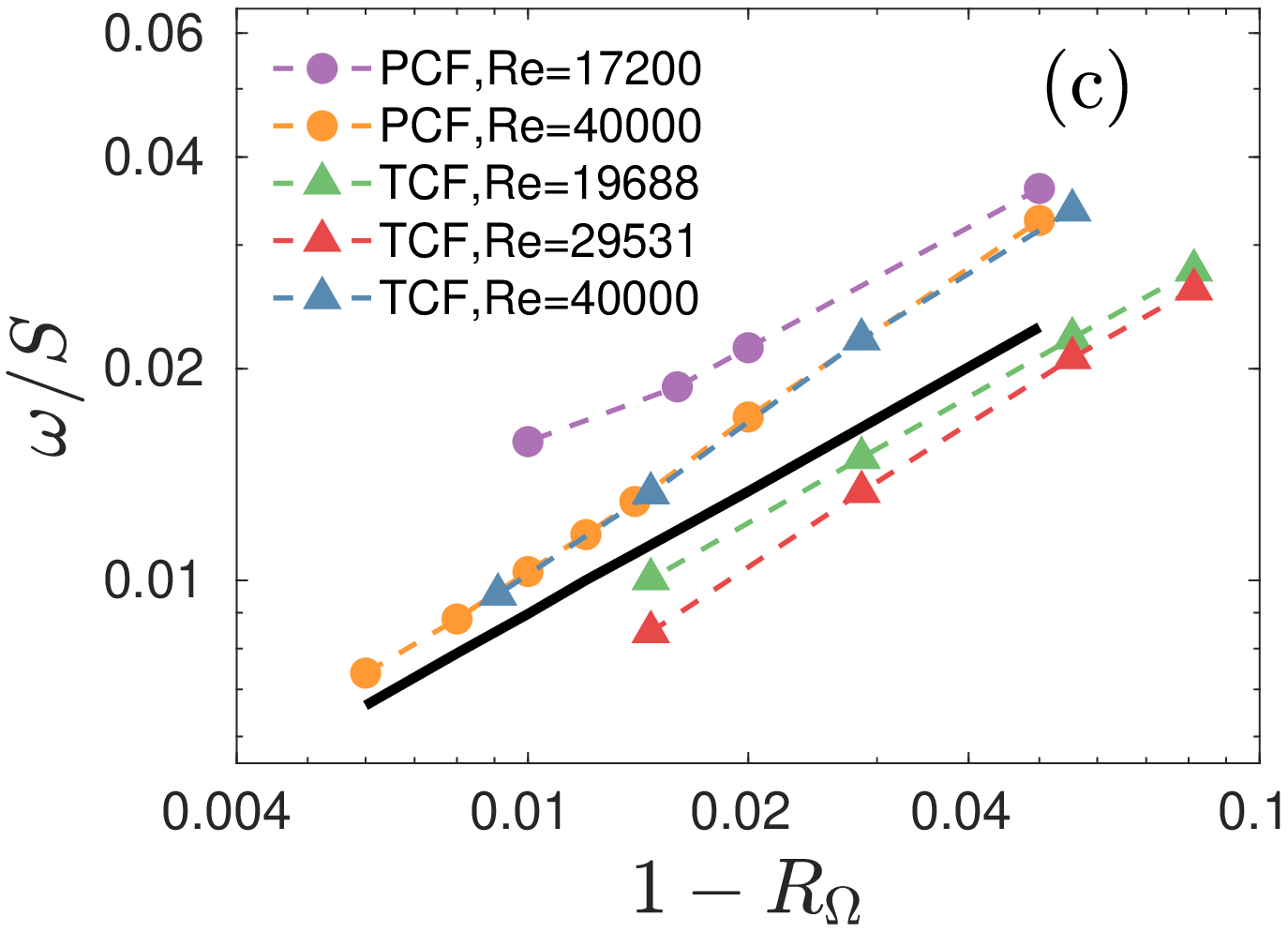}
\end{center}
\caption{({\it a}) Time series of $K$ in PCF at $\Ro=0.97$ 
and TCF at $\Ro=0.98$ and $\Rey=40\,000$.
Time is non-dimensionalized by the shear rate $S=\Delta U/d$.
({\it b}) Phase space plot of volume integrated $K$ and enstrophy $\omega^2$, and 
wall shear stress $\tau_w$ in PCF at $\Rey=40\,000$ and $\Ro=0.99$.
Yellow and blue colours indicate large and small
temperature fluctuations, respectively.
({\it c}) Frequency of the bursts non-dimensionalized by $S$ for PCF and TCF. The
black solid line gives the growth rate of 
the most unstable mode in PCF predicted by linear theory.
\label{time}}
\end{figure}

Couette flows are fully turbulent at higher $\Rey$,
also when HTE is maximal, leading to plume-like thermal structures 
(figure \ref{vis}{\it c},{\it d}).
Above $\Rey\approx 10^4$ high values of HTE
are accompanied by strong recurring low-frequency
bursts of turbulence in both PCF and TCF.
Such turbulent bursts are evident in time series of the 
volume integrated turbulent kinetic energy $K$ (figure \ref{time}{\it a})
and emerge if $\Ro \gtrsim 0.94$.
Those are persistent and approximately periodic
and come along with bursts of enstrophy and temperature fluctuations
and significant changes of shear stresses and heat fluxes at the wall. 
A phase-space plot illustrates the approximate limit cycle dynamics (figure \ref{time}{\it b}).
The burst frequency declines for $\Ro \rightarrow 1$ because the flow becomes more stable
and follows a similar scaling in all cases (figure \ref{time}{\it c}).
During the approximate limit cycle oscillations PCF and TCF is supercritical and
continuously though weakly turbulent between the bursts. 
I have verified that PCF is also linearly unstable
if the mean velocity profile from the DNSs instead of the laminar
one is used in the stability analysis.
The growth rate of the most unstable mode follows a similar trend
as the burst frequency (figure \ref{time}{\it c}), suggesting
that the bursts are related to linear instabilities.

\section{Concluding remarks}

The key conclusion of my study is that 
even a simple Coriolis body force can strongly change
heat/mass transfer rates and can make 
heat/mass transfer much faster than momentum transfer
in shear flows,
as indicated theoretically recently \citep{Alben,Motoki}.
Optimization of heat/mass transfer by body forces is thus a
promising avenue for further research.
The mechanism of momentum transport reduction 
by the Coriolis force does not depend on $\Rey$, 
implying that the observed dissimilarity between momentum and 
heat/mass transfer,
found in both plane Couette and Taylor-Couette flow, persists at higher $\Rey$.
The highest dissimilarity happens in rotating Couette flows close to
the inviscid neutral stability state when momentum transfer is stronger
reduced than heat/mass transfer. 
Also other rotating shear flows tend to evolve towards the neutral stability state 
\citep{Metais,Barri}, suggesting that heat and mass 
are transported much faster than momentum in such flows.
Dissimilarity between momentum and heat transport is also found in
rotating channel flow \citep{Matsubara,Geert2018,Geert2019},
albeit in a limited region where the flow approaches the zero-absolute-mean-vorticity state, 
and in shear flows with buoyancy forces \citep{Li,Pirozzoli2017}.
In DNS and rapid distortion theory of
rotating uniformly sheared turbulence \cite{Geert2005} observed turbulent
Prandtl numbers much smaller than one when the zero-absolute-mean-vorticity
state is approached, which also implies fast heat transport.
This all suggests that more engineering and astrophysical flows 
display dissimilarities between heat or mass transfer and momentum transfer.
Another implication of the present study is that
heat and mass transfer modelling in flows with body forces
requires careful considerations since the Reynolds analogy can fail.

\section*{Acknowledgements}
SNIC is acknowledged for providing 
computational resources in Sweden.
The author further acknowledges financial support
from the Swedish Research Council (grant number 621-2016-03533).
The author is grateful to Prof. G. Kawahara and Dr. S. Motoki
for the data on optimal transport in PCF, and to Bendiks Jan 
Boersma for providing and helping with the DNS code for TCF.

\section*{Declaration of interests}
The author reports no conflict of interests

\bibliographystyle{jfm}
\bibliography{ref.bib}

\end{document}